\documentclass[12pt]{article} 
\usepackage{url,hyperref,lineno,microtype,subcaption,amssymb,amsmath}
\usepackage[margin=0.7in]{geometry}
\usepackage[onehalfspacing]{setspace}
\usepackage{graphicx}
\usepackage[english]{babel}

\begin{document}

\title{Some notes about the current researches on the physics of relativistic jets}
\author{Luigi Foschini\footnote{Brera Astronomical Observatory, National Institute of Astrophysics (INAF), 23807 Merate (LC), Italy. Email: \texttt{luigi.foschini@inaf.it}}}
\date{November 29, 2021}
\maketitle

\begin{abstract}
Some highlights of the recent researches in the field of relativistic jets are reviewed and critically analyzed. Given the extent of the available literature, this essay symbolically takes the baton from the outstanding and recent review by Blandford, Meier, and Readhead (2019). Therefore, I focus mostly on the results published during the latest few years, with specific reference to jets from active galactic nuclei.  
\end{abstract}

\section{Introduction}
It is quite challenging to write a review on relativistic jets just a couple of years after the publication of the excellent essay written by giants like Roger Blandford, David Meier, and Anthony Readhead \cite{BLANDFORD2019}. It would be sufficient to look at their publication list to have an almost complete history of our understanding of relativistic jets. Therefore, on one side, I will necessarily refer to their work to recall most of basic concepts; on the other side, I expand some topics and add new information, mostly on the basis of the works published after the above cited review. As already noted by Blandford, Meier, and Readhead \cite{BLANDFORD2019}, this research field is today so large that it is impossible to write an exhaustive review. Nevertheless, one can try to write at least a self-consistent and complete essay, with many apologies to those who are not cited. Therefore, rather than a review, this is a personal selection of topics in the field of relativistic jets. This means that the present essay is mostly focused on jets from Active Galactic Nuclei (AGN), with some sporadic reference also to X-ray Binaries (XRBs). I do not deal with Gamma-Ray Bursts (ultrarelativistic jets) and protostars (supersonic jets). 

To summarize the basic results emphasized in the Blandford et al.'s review \cite{BLANDFORD2019}, a relativistic jet is a collimated outflow of particles, with bulk velocity close to that of light, extending from the poles of a rotating and accreting compact object, which can be either a black hole or a neutron star. The magnetic field necessary for the collimation and acceleration of particles is generated by the accretion disk and advected close to the horizon. The shape of the jets is self-similar (parabolic, conic). The most powerful jets can keep the collimation for hundreds of parsecs for AGN and a few thousandths of parsecs for XRBs, ending in strong hot spots, while the low-power ones are quickly degraded into wide plumes. Jets display a wide variety of sub-structures depending on the interaction with the environment, such as standing shocks and bending. The electromagnetic emission covers all the measurable spectrum, and it is speculated that these structures may significantly contribute to the cosmic rays spectrum, although the observations supporting this idea are still rather weak and indirect\footnote{This is mostly a problem of detecting cosmic rays and identifying the sources (see, for example, \cite{AUGER} for a review on the latest results of the Pierre Auger Observatory). Charged particles are deviated by the Galactic and local magnetic fields, so that only extremely energetic particles (negligible deviation) can be useful for astronomy; however, these are relatively rare and the position error is still large (degrees). Neutrinos are difficult to detect, and neutrons have too short a lifetime. The association of relativistic jets with neutrinos seems to be the most promising research field \cite{PLAVIN2020,PLAVIN2021,HOVATTA2021}.}. Jets were more common in the early universe, because these structures are an efficient relief valve for the angular momentum, thus allowing the high accretion rates of young quasars. Their influence can extend far beyond the central parsec of the AGN, affecting the whole host galaxy and the intergalactic medium. 

\begin{table}[ht]
\caption{How basic questions/aims on astrophysical jets from AGN have changed after about forty years of studies. From \cite{BLANDFORD2019,FANTI1983}. There is no one-to-one correlation between the two lists. The questions/aims are just in the same order as they were written by the authors in their papers.}
\footnotesize
\begin{center}
\begin{tabular}{p{1cm}p{7cm}p{7cm}}
\hline
\# & {\bf Fanti (1982)} & {\bf Blandford, Meier, Readhead (2019)}\\
\hline
1 & Velocity in jets: supersonic? relativistic? constant? & Discover new information about gas flow and jet production around black holes using millimeter and submillimeter EHT observations\\
2 & Energy carried by jets: which forms? which transformations? & Understand if the sustenance of a magnetic field near a black hole, which determines whether an AGN is radio loud or radio quiet, is due to physical processes near the black hole or controlled by the infalling gas\\
3 & Mass carried by the jets: where does it come from? & Employ VHE neutrino observations to open up a new window on AGNs and determine if blazar jets are major cosmic ray sources\\
4 & Jets collimation: free jets? confined? by what? & Use EHT to study many jets on the scale of $R_{\rm inf}$ and verify that the FR class of a radio source is determined here\\
5 & Structure of magnetic field: helical? tangled? & Learn how to map emission in a given spectral band onto jet radius\\
6 & Jets heavy or light? & Determine if the TeV emission from blazars is synchrotron or Compton radiation\\
7 & Origin of wiggles: gravitational interaction? precession? K-H instabilities? & Develop hybrid numerical simulations that meld relativistic hydromagnetic descriptions with particle kinetics and radiative transfer\\
8 & Why are there one-sided jets? & Quantify the role of AGN jets in promoting and limiting galaxy formation and evolution\\
\hline
\end{tabular}
\end{center}
\normalsize
\label{questions}
\end{table}%

While reading papers to prepare this review, I came upon an old volume, the proceedings of the international workshop {\em Astrophysical Jets}, held in Torino (Italy), on October 7-9, 1982. I found an interesting essay by Roberto Fanti \cite{FANTI1983}, where he listed a series of open questions about relativistic jets in AGN (see Table~\ref{questions}). It is interesting to compare that list with the similar one proposed by Blandford, Meier, and Readhead \cite{BLANDFORD2019}, because it shows how our knowledge on astrophysical jets improved during the last forty years. Most of Fanti's questions have now reasonable answers, which can be read in the excellent review by Blandford, Meier, and Readhead \cite{BLANDFORD2019}. However, the availability of new instruments might refresh old questions: for example, the forthcoming launch of the \emph{Imaging X-ray Polarimetry Explorer (IXPE)}\footnote{\url{https://ixpe.msfc.nasa.gov/}.}, scheduled on December 9, 2021, might reopen the question 5 of the Fanti's list. Obviously, no question is closed forever in science, but it is simply abandoned when there is no more interest to search for alternatives. 

\section{High-resolution radio observations}
Surely, the greatest most recent advancements in understanding the jet physics were due to the development of very high resolution radio observations (see, for example, \cite{HADA2019} for a detailed review on this topic). Most of us will permanently remember the amazing images of the photon ring around the supermassive black hole in M87 ($z=0.00428$) generated with the data obtained by the {\em Event Horizon Telescope} (EHT), which confirmed the expectations from the main theories \cite{EHT2019I,EHT2019II,EHT2019III,EHT2019IV,EHT2019V,EHT2019VI}. Being the first EHT observation it had a relatively limited impact on our knowledge of the physics of relativistic jets\footnote{A second paper on the jet of Centaurus A ($z=0.00183$) was published this year \cite{EHT2021IX}. Although EHT successfully imaged the jet region close to the black hole, it cannot image its shadow. The singularity of Cen A is $\sim 5.5\times 10^{7}M_{\odot}$ and, therefore, its shadow should be observable with EHT at THz frequencies, according to the scaling laws \cite{HEINZ}.}, but the few results referred to one source are nonetheless of great importance. EHT confirmed the presence of a rotating supermassive black hole in M87, with $M\sim 6.5\times 10^{9}M_{\odot}$ \cite{EHT2019VI}, consistent with the stellar dynamics measurements, but not with the gas dynamics ones. A possible explanation has already been presented, by invoking a thick accretion disk with angular momentum decoupled from that of the black hole (misalignment $>11^{\circ}$, typical $\sim 27^{\circ}$, \cite{JETER2021}). Comparison with numerical models confirmed that the jet is powered by the rotating energy of the spacetime singularity, but with no significant preference between the adopted models \cite{EHT2019V}. Later in 2021, the analysis of EHT polarimetric data favored a magnetically-arrested accretion disk model (MAD), characterized by a strong magnetic field ($B\sim 1-30$~G as measured by EHT) advected close to the event horizon \cite{EHT2021VII,EHT2021VIII}. Significant simplifications were adopted to obtain these results and a detailed critical discussion was presented by Gralla \cite{GRALLA2021}. On a more general point of view, it is worth citing the Very Long Baseline Array (VLBA) study on III~Zw~2 ($z=0.0898$), a $\gamma-$ray intermediate Seyfert \cite{LIAO2016}, which revealed that the measured magnetic field in the innermost region was well below the value prescribed by the MAD model \cite{CHAMANI2021}. III~Zw~2 has an accretion rate greater than M87 ($\sim 0.04$ vs $\sim 2\times 10^{-5}$, \cite{BERTON2015,EHT2019V}), which implies a likely different accretion disk, much more radiatively efficient. Numerical models such as MAD \cite{IGUMENSHCHEV,NARAYAN2003} are suitable for radiatively-inefficient disks and, therefore -- unsurprisingly -- cannot fit to more efficient disks. But this underlines the need to couple General Relativistic MagnetoHydroDynamic (GRMHD) numerical models with radiative transfer models to fully explore all the types of disk-jet coupling (this is the question 7 in the Blandford et al.'s list, see Table~\ref{questions}).

I just would like to add one more assumption to be tested in the forthcoming EHT observations. All the models adopted by EHT assumed that the angular velocity of the magnetic field threading the event horizon ($\omega_{\rm F}$) is half the value of the black hole one ($\omega_{\rm F}=0.5\omega_{\rm H}$). This is a common assumption in almost all the models of jets, because it corresponds to the maximum efficiency in the jet production according to the Blandford-Znajek theory (BZ, \cite{BZ}). This was never tested, because of the obvious problem of measuring the magnetic field close to the horizon\footnote{Also measuring the spin of a black hole is not so easy, but at least there are different methods continuously improving (see \cite{BAMBI2021} for a recent review).}. But now, EHT observations could fill this gap. In addition, measuring the interplay (if any) between $\omega_{\rm F}$ and $\omega_{\rm H}$, might open a major understanding of the jet generation, and the differences between jetted and non-jetted AGN \cite{FOSCHINITEX} (this is also the question 2 in the Blandford et al.'s list, see Table~\ref{questions}). This measurement can also verify the hypothesis by David Garofalo, according to whom retrograde disks can power jets better than prograde ones \cite{GAROFALO2010,GAROFALO2017}. 

It is worth noting that the BZ theory (published in 1977, \cite{BZ}) is still triggering lively and interesting exchange of views: in Summer 2021, King and Pringle \cite{KING2021} challenged the theory, claiming that the accretion of electric charges onto the singularity horizon can screen the electric field, thus invalidating the BZ mechanism. Komissarov \cite{KOMISSAROV2021} replied, proving that they were wrong: although the charge accumulation can change the dynamics of the ergosphere, it cannot affect the gravitationally induced electric field. A nice and pleasant review on numerical simulations of jets, including the story of more than forty years of debates on the BZ theory, can be read in \cite{KOMISSAROV2021B}.

Another important advancement was done by the {\em LOw Frequency ARray} (LOFAR\footnote{\url{https://www.astron.nl/telescopes/lofar/}}), which increased the sensitivity (fraction of mJy) and the angular resolution (subarcsecond) in the MHz frequency range ($30-240$~MHz), coupled with a very large field of view (up to $1200$~square degrees) \cite{HAARLEM2013}. Among the recent and interesting results, I would like to mention the study of the evolution of jet bubbles in a group of galaxies and the interaction with the intergalactic medium \cite{BRIENZA2021} and the challenge to the Fanaroff-Riley paradigm, according to which the radio morphology depends on the observed radio power: also FRI-type radio galaxies can generate jets, which remain collimated for long distances and form hot spots \cite{MINGO2019}. 

\section{Scaling laws}
\label{scalaws}
As the angular resolution of radio telescopes increased, it became possible to study the jet shape from the large scale down to the innermost region, close to the central black hole. Obviously, the best studied AGN is M87 \cite{NAKAMURA}, followed by 1H~$0323+342$ ($z=0.063$) \cite{HADA2018,DOI2018}, and a significant sample of nearby jetted AGN (e.g. \cite{KOVALEV2020,BOCCARDI2021}). All these studies confirmed the self-similarity of the jet shape: the innermost part has a parabolic shape, then, after a collimation break, the shape changes to conical. 

The self-similarity is a critical hypothesis for the application of the scaling laws developed by Heinz \& Sunyaev \cite{HEINZ}. According to these laws, the jet power depends on the mass of the central black hole and the type of the accretion disk (whether dominated by advection, gas-, or radiation-pressure). These functions are not linear: for example, in the case of a radiation-pressure dominated disk, the jet power $P_{\rm jet}$ is proportional to the mass of the central compact object $M^{17/12}$, while in the other cases (advection- and gas-dominated disk) there is also a component as a function of the accretion rate (see Table~1 in \cite{HEINZ}). 

These laws were immediately applied to generate the so-called ``fundamental plane'' \cite{MERLONI2003,FALCKE2004}, which correlates the luminosities at $5$~GHz (proxy of the jet power) to those in the $2-10$~keV energy band (proxy of the accretion luminosity), together with the mass of the central black hole. Starting from measured quantities is quite a simple approach, but it suffers the well-known drawback ``correlation is not causation''. The $2-10$~keV energy band can be dominated by either the accretion disk corona in XRBs and non-jetted AGN, or the inverse-Compton emission of the jet in flat-spectrum radio quasars (FSRQs) and jetted narrow-line Seyfert 1 galaxies (NLS1s), or the jet synchrotron emission in high-frequency peaked BL Lac Objects. Also the 5~GHz frequency is sufficiently low to suffer contamination in the case of strong starburst activity (e.g. \cite{KLEIN2018}). 

Merloni et al. \cite{MERLONI2003} stated that they want to study the disk-jet connection, but then, to avoid the above cited problems for the X-ray measurements, they did not consider jetted AGN (sic!). Therefore, the proposed disk-jet connection might make some sense for XRBs, but it is meaningless for AGN, because the considered sample is composed of non-jetted AGN (with a few exceptions). The radio detection is not sufficient to claim for the presence of a jet, because it might come from different processes in the accretion disk (synchrotron emission, winds,... cf \cite{ABRAMOWICZ,PANESSA2019}), or due to star formation. The simple fit to the correlation does not justify the hypothesis of a jet: as clearly shown by Marco Chiaberge, also the Sun, the Moon, Jupiter, and Saturn fit well the correlation, but surely they do not have jets (see Fig.~1 in \cite{CHIABERGE2007}).

Falcke et al. \cite{FALCKE2004} did consider jetted AGN (BL Lacs and FRI radio galaxies), and mix them with Low-Luminosity AGN (LLAGN), Sgr~A*, and XRBs. The same doubt raised by Chiaberge \cite{CHIABERGE2007} on the Merloni's plane is still valid here: correlation is not causation, again. In addition, the origin of the radio emission in LLAGN is still matter of debate, with articles supporting the jet emission and others favoring the star formation (e.g. \cite{CHIARALUCE2019,ALGERA2020,SILPA2020,WEBSTER2021,RADCLIFFE2021}). Sgr~A* generated a relativistic jet in the remote past (Myrs ago, e.g. \cite{CECIL2021}), but surely not today (e.g. \cite{GOLDWURM2007,PONTI2017}). Even if the radio and X-ray emission of LLAGN and Sgr~A* is due to a jet, then the resulting plane might be a visualization of the self-similarity of the jet emission at radio and X-rays. However, this means -- again -- that the plane is no more probing the disk-jet coupling. Many other works underlined these and other flaws (see \cite{CHIABERGE2007,FOSCHINI2014,SAIKIA2018,BERTON2019,FISCHER2021}, just to cite a few). Nevertheless, these types of fundamental planes are still luring researchers, because of their ease. 

Another example is the unified model of disk-jet coupling in stellar-mass black holes \cite{FENDER2004,MCCLINTOCK2006}, where there is a mixture of physical quantities (such as the luminosity of the accretion disk calculated by adopting simple models) with simple raw measurements (such as a hardness ratio built with the detector counts of the \emph{Rossi X-ray Timing Explorer} (RXTE) in the $6.3–10.5$~keV and $3.8–6.3$~keV energy bands). Also the use of qualitative terms such as soft and hard can generate significant misunderstandings, because they depends on the energy band considered. 

Back to scaling laws, it is worth stressing that Heinz \& Sunyaev's \cite{HEINZ} theory really apply to physical quantities: jet luminosity, mass of the central black hole, accretion rate. There are plenty of publications with simple and approximate equations to calculate these quantities from observations. So, why not use them? Don't be lured by the myth of model-independent measurements: it is exactly a myth\footnote{Heinz \& Sunyaev \cite{HEINZ} stated that their laws are model independent, but they fail to highlight that their starting assumptions are the self-similarity of the jet shape and the synchrotron emission as the dominant radiative process.}. The scaling laws are still confirmed by using derived physical quantities (\cite{FOSCHINI2014,FOSCHINI2017}), so that one has no more need to clutch at straws by cherry picking the right sample to make the fundamental plane work. In addition, since these scaling laws refer to the relativistic jet, the central compact object does not necessarily need to be a black hole. Jets from accreting neutron stars can also fit well in the Jet Power-Disk Luminosity Plane ($P_{\rm jet}-L_{\rm disk}$, JD-plane, \cite{FOSCHINI2014,FOSCHINI2017}) forming the low-mass branch of the XRBs. The corresponding branch of the AGN population is represented by NLS1s. Therefore, the mass of the central compact object is the main driver of the unification, not the states of the accretion disk (cf \cite{FALCKE2004,YUAN2014}). 

The inclusion of NLS1s in the population of jetted AGN had also another critical effect: the break down of the blazar sequence, that well-known trend linking FSRQs (highest jet power, lowest frequency of the synchrotron peak) to BL Lac Objects (lowest jet power, highest frequency of the synchrotron peak) \cite{FOSSATI1998}. This trend is explained in terms of radiative cooling of relativistic electrons \cite{GHISELLINI1998}: FSRQs have strong accretion disks and a photon-rich environment, which implies an efficient dissipation of energy and high jet power; BL Lac Objects have weak disks and a photon-starving environment, resulting in an inefficient dissipation and low jet power. It is worth noting that the masses of the central black holes of the two sub-classes are more or less of the same order of magnitude. The discovery of high-energy gamma rays from NLS1s \cite{LAT1,LAT2,LAT3,FOSCHINI2010} added a new class of powerful jetted AGN. NLS1s have observational characteristics similar to FSRQs (strong accretion disk, photon-rich environment), but their masses of the central black hole and jet powers are smaller. This can be explained by invoking the Heinz \& Sunyaev's \cite{HEINZ} scaling laws ($P_{\rm jet}\propto M^{17/12}$) and taking into account that the mass of the central black hole in NLS1s is smaller than that of FSRQs ($10^{6-8}M_{\odot}$ vs $10^{8-9}M_{\odot}$, e.g. \cite{FOSCHINI2015}). Therefore, the blazar sequence (jet power as a function of electron cooling) holds only for objects with similar masses, but it breaks down when significantly different masses are considered (see Fig.~1A in \cite{FOSCHINI2017}). This is evident when unifying XRBs and AGN, because a linear scaling with the mass is not sufficient (cf \cite{FOSCHINI2011}). 

The blazar sequence has also been challenged on the basis of the synchrotron peak frequency vs peak luminosity plane ($\nu_{\rm peak}-L_{\rm peak}$) of many more jetted AGN ($15\times$ the number of sources used by \cite{FOSSATI1998}) of different classes, resulting in the confirmation of the selection bias \cite{KEENAN2021}. 

More details on the impact of the discovery of powerful relativistic jets from NLS1s can be found in these recent reviews \cite{FOSCHINI2012,FOSCHINI2020}.

\section{Where does the jet dissipate most of its energy?}
Larger programs based on VLBA and Very Long Baseline Interferometer (VLBI) radio observations have had a much larger impact in developing the physics of jets during the latest decades: VLBA at $43$ and $86$~GHz program of the Boston University Blazar Group\footnote{\url{https://www.bu.edu/blazars/VLBAproject.html}.}, the MOJAVE Program\footnote{{\bf M}onitoring {\bf O}f {\bf J}ets in {\bf A}ctive galactic nuclei with {\bf V}LBA {\bf E}xperiments, \url{https://www.physics.purdue.edu/MOJAVE/}.} at $15$~GHz, TANAMI\footnote{{\bf T}racking {\bf A}ctive galactic {\bf N}uclei with {\bf A}ustral {\bf M}illiarcsecond {\bf I}nterferometry, \url{https://pulsar.sternwarte.uni-erlangen.de/tanami/}.} at $8.4$ and $22$~GHz, all with images at milliarcsecond resolution. Making the data publicly available increased the impact of these programs outside their respective core groups. An invaluable support also came from large monitoring programs based on single dish observations, such as Mets\"ahovi\footnote{\url{http://www.metsahovi.fi/opendata/}.} ($37$~GHz, about $40$ years observations!), the University of Michigan Radio Astronomy Observatory\footnote{\url{https://dept.astro.lsa.umich.edu/datasets/umrao.php}} at $4.8$, $8$, and $14.5$ GHz, Owens Valley Radio Observatory\footnote{\url{https://sites.astro.caltech.edu/ovroblazars/index.php?page=home}.} (OVRO, $15$~GHz), and Effelsberg\footnote{\url{https://www3.mpifr-bonn.mpg.de/div/vlbi/fgamma/fgamma.html}.} (F-GAMMA, several frequencies, from $2.64$ to $42$~GHz).

The programs based on high-resolution imaging observations allowed to observe and to monitor for long time the evolution of jet structures and components, to directly measure the apparent speed of the latter ($\beta_{\rm app}$), the brightness temperature, linear and circular polarization, Doppler and Lorentz factors, jet opening angles, and many other quantities, for hundreds of jetted AGN (e.g. the MOJAVE program \cite{KHARB2010,LISTER2016,LISTER2019,LISTER2021,HOMAN2021}). This is an incredible wealth of data showing the amazing complexity of relativistic jets. Coupling these observations with measurements at other, much higher, frequencies (optical, X-rays, gamma rays), has opened intriguing perspectives. Already during 1990s, coupling VLBA observations with the data of the EGRET instrument onboard the \emph{Compton Gamma-Ray Observatory (CGRO)} resulted in the association -- in some cases -- of gamma-ray emission and the passage of a plasma blob through the millimeter radio core or other features downstream, located at parsec scales (e.g. \cite{JORSTAD2001}). This cast doubts on the mainstream theory that most of the dissipation occurs in the innermost region of the AGN, around the broad-line region (BLR, $r\sim 10^{3}r_{\rm g}$, where $r_{\rm g}=GM/c^2$ is the gravitational radius, e.g. \cite{GHISELLINI2010}). This result was recently confirmed by correlating VLBA observations with the data of the LAT instrument onboard the \emph{Fermi} satellite (e.g. \cite{JORSTAD2017}; see also the interesting results by \cite{PLAVIN2019} obtained by coupling the high positional resolution of VLBI and \emph{Gaia} satellite). 

Both theories suffer drawbacks. Against the BLR hypothesis there is the lack of observations of absorption features in the gamma-ray spectra \cite{COSTAMANTE2018}, but the current operational instruments might not be best suited to accomplish this task. The final word may be at hands with the forthcoming \v{C}erenkov Telescope Array (CTA) \cite{ROMANO2020}. The dissipation at parsec scale is even more difficult: since the jet shape is self-similar, and it is assumed that the perturbations start from the innermost region around the central black hole, the size of the emitting region increases with distance from the black hole. Therefore, when the perturbation arrives at distances of the order of one parsec, its size can be too large to reconcile with the observed very short (minutes) variability derived from $\gamma-$ray data analysis \cite{MEYER2019} or from the exceedingly high brightness temperature \cite{JORSTAD2017,KOVALEV2016,GOMEZ2016,PILIPENKO2018}, unless one assumes an unrealistically large Doppler factor. The relativistic magnetic reconnection process offers a viable solution (e.g. \cite{PETROPOLOU2016}): in this case, the size of the emitting region is detached from its distance from the central black hole and could be as small as necessary with common values of the Doppler factor. Other alternative scenarios have been proposed, such as structured jets (spine+layer, \cite{GHISELLINI2005,PINER2018}), jet-in-jet (based on magnetic reconnection too, \cite{GIANNIOS2009}), or deviation from equipartition \cite{PETROPOLOU2015,NOKHRINA2017}. In the case of the brightness temperature, the possibility that the current technology is not yet able to resolve the real emitting region is also quite reasonable. One has also to take into account that the real question is not about the one or the other dissipation region, but it can be more likely that the same source changes its dissipation region (from BLR to pc-scale and vice versa) depending on some yet unknown trigger \cite{FOSCHINI2011B,GHISELLINI2013}. The polarimetry in the X-ray energy band that will be provided by \emph{IXPE} might help to solve some of these questions, at least in the case of high-frequency peaked BL Lac Objects \cite{MARSCHER2021,TAVECCHIO2021}. 

Among the different intriguing cases, there is 1H~$0323+342$, because it was possible to build a detailed map of the AGN and the central region of the host galaxy \cite{FOSCHINI2019}. There is a nozzle spotted with VLBA observations of the MOJAVE program \cite{DOI2018}, which is quite far from the singularity, about one order of magnitude farther than the dusty torus ($\sim 54$~pc vs $5.7$~pc, \cite{FOSCHINI2019}), but still within the narrow-line region (NLR). Akihiro Doi \cite{DOI2018} suggested the possibility of a correlation with high-energy gamma-ray activity studied by Paliya \cite{PALIYA2014}, but the sparseness of data prevented to set significant constraints. Nonetheless, the hypothesis of a strong dissipation at the level of the NLR is challenging. The interaction of jets and winds with the NLR has been already observed by studying the [OIII] emission lines profiles \cite{KOMOSSA2008,BERTON2016B,KOMOSSA2018}, and could be the gate for the AGN/host galaxy feedback \cite{BERTON2021}. The Multi Unit Spectroscopic Explorer (MUSE) of the Very Large Telescope (VLT) at the European Southern Observatory (ESO) is the best instrument to explore this feedback. Recent surveys revealed that FRI radio galaxies have compact optical emission line regions ($\sim 1$~kpc), while more powerful FRII radio galaxies are associated with larger regions ($\sim 17-18$~kpc, up to $80$~kpc) \cite{BALMAVERDE2018,BALMAVERDE2019,BALMAVERDE2021} (this is also the question 8 in the Blandford et al.'s list, see Table~\ref{questions}).

Coupling high-resolution imaging at different frequencies (from radio to optical, even X-rays) will surely be the cornerstone for the forthcoming studies. Particularly, the project of X-ray interferometry ($20~\mu$as at $6$~keV) can represent the hinge of a revolution in understanding relativistic jets \cite{UTTLEY2021}.  

\section{Emergent topics}
\subsection{Absorbed or Intermittent Jets?}
Another gem from one of these large monitoring programs, specifically the Mets\"ahovi Radio Observatory of the Aalto University (Finland), is the discovery of relativistic jets from radio quiet/silent AGN, specifically NLS1s \cite{LAHTEENMAKI2018}. The radio telescope is a single dish operating at $37$~GHz and, between $2014$ and $2018$, detected sporadic, but intense (at Jansky level) emission from a sample of sources classified as radio quiet or even silent. Such intense flux density at such high frequency can be due only to a relativistic jet, but the source was classified as radio-quiet/silent implying no jet. Is the classification not correct or is it another type of changing-look AGN? Marco Berton and Emilia J\"arvel\"a proposed that the jet could be absorbed at radio frequencies, because either synchrotron self-absorption, or free-free absorption due to a nearby plasma shield (starburst activity or shocks) \cite{BERTON2020,JARVELA2021}. Since the classification radio-quiet/silent/loud is done by taking the flux density at $5$ or $1.4$~GHz, where the free-free absorption is expected to act, this implies that the quietness or loudness as a proxy of the jet presence is meaningless (see also Sect~4 in \cite{BERTON2021BIS}). Preliminary results of observations at $15$~GHz and at X-rays confirm the emergence of the jet emission at higher frequencies\footnote{See the slides of M. Berton's seminar, held on October 28, 2021: \url{http://www.brera.inaf.it/interroga/dbServer?cmd=foto&oid=66119465&formato=pdf&database=web&options=Absorbed+jets+in+narrow-line+Seyfert+1+galaxies}}. Therefore, this is just one more step to overcome the old concept of radio quiet/loud/silent, and to change the classification to a more physical jetted/non-jetted AGN, as proposed by Paolo Padovani \cite{PADOVANI2017}.   

There is another intriguing case, TXS~$0128+554$ ($z=0.0365$), a baby (kinematic age $82\pm 17$~years) Compact Symmetric Object (CSO) generating a powerful (detected at high-energy gamma rays), but intermittent jet \cite{LISTER2020} (another similar case is PKS~$1718-649$, $z=0.014$, kinematic age $\sim 100$~years, \cite{MIGLIORI2016}). Episodic activity might be a characteristic of young AGN. These objects are likely to be born in a photon and matter rich environment, and the jet has to fight a battle against the surrounding matter to develop. This generates the context where the jet can be intermittent, because the accretion disk can be clogged by too much accreting matter, or it can be absorbed, or stopped, or bent by the surrounding plasma. These young AGN offer us a unique, but underrated, opportunity: given the small redshift, it is easy to observe them, but we can gain invaluable knowledge on how AGN were born. 

\subsection{``Minion'' radio galaxies}
During the last decade, another population of jetted AGN emerged: miniature radio galaxies, with almost no extended radio emission and low jet power, also known as FR0, to be distinguished from the classical classification FRI/FRII from Fanaroff and Riley \cite{BALDI2009,BALDI2018}. The lack of extended radio emission might be interpreted as the indication of a young source evolving to the classical FRI/FRII types, but the observational properties suggest a different explanation, either a jet with insufficient power to generate extended radio emission or an intermittent activity \cite{BALDI2018,TORRESI2018}. High-energy gamma-ray emission has been detected from a few FR0 \cite{GRANDI2016,PALIYA2021}, suggesting that this class of sources might be a suitable candidate for neutrino emission \cite{TAVECCHIO2018}. However, the jet in FR0 seems to be too weak to efficiently generate neutrinos, and it has been suggested that protons can escape from the jet and interact with the interstellar gas of the host galaxy \cite{TAVECCHIO2018}, a mechanism similar to what has been advanced for starburst galaxies or when there is also a contribution from outflows \cite{LAMASTRA2016}. Although none explored this hypothesis yet, it would be interesting to apply this mechanism to NLS1s, where jet, outflows, and starburst activity are coexistent \cite{CACCIANIGA,LONGINOTTI2015,GIROLETTI2017}. 


\subsection{Gravitational waves and Tidal Disruption Events}
The recent opening of the gravitational astronomy had also an impact on the study of relativistic jets. The current facilities can observe only stellar-mass black holes/neutron stars: the merging generates a remnant, which collapses and triggered a GRB jet (e.g. \cite{MURGUIA2021}). The possibility to detect in the forthcoming years\footnote{\url{https://www.elisascience.org/}.} gravitational waves also from supermassive black holes refreshed studies and monitoring of candidates binary jetted AGN (e.g. OJ287 \cite{KOMOSSA2021}) and of recent mergers \cite{CHIABERGE2017,CHIABERGE2018}. Merging events are of particular interest for relativistic jets, because it is often suggested a link between the two phenomena (e.g. \cite{BERTON2019B,PALIYA2020,JARVELA2020}). While on stellar-mass compact object the onset of the jet is due to the collapse of the remnant, the mechanism at action in the merging of two supermassive black holes must be quite different, likely due to the increase of available gas for the accretion. But this will be better investigated when forthcoming facilities will make it possible the detection of gravitational waves from binary AGN.

A similar topic is that of the Tidal Disruption Events (TDEs), which attracted significant interest during the last decade \cite{DECOLLE2020}. In this case, a star is orbiting so close to a black hole to be disrupted by tides. Again, the increase of the accretion onto the black hole can trigger the formation of a jet as a relief valve of the angular momentum in excess \cite{BURROWS2011}. It has also been suggested that certain gamma-ray flares, and particularly some episodes of variability on short timescales, might be explained as the passage of a star through the jet cone \cite{BARKOV2012,KHAN2013}.

\section{Final remarks}
I have no list of questions or aims for the conclusions of the present essay. Those presented in Table~\ref{questions} are more than sufficient and challenging. A recent work we are doing suggests me to emphasize the research in the available literature, and to avoid relying too much on automatic classifications and cross-matching between catalogs. We recently performed a reanalysis of a part (2982 sources) of the 4th {\em Fermi} LAT Point Source Catalog \cite{4FGL}, by checking all the available literature for each source: it resulted that many sources needed to be reclassified, with a significant increase of jetted NLS1s/Seyferts, and misaligned AGN. We also found many reports on sources displaying strong changes in the optical spectra, from line-dominated spectra to featureless ones (changing-look AGN). We also found that many of the commonly adopted redshifts were not correct, because of mere transcription errors from the paper to the database, old and outdated information, lower limits taken as measurements, and many other problems \cite{FOSCHINI2021}. 

This is really a time-consuming work, but nonetheless it must be done, otherwise, we will continue to study and to draw fake conclusions on strongly biased samples. It is necessary to find new working solutions to take into account the treasure of the published literature. I do not think that this task can be done by a computer, even if based on trained neural networks. Perhaps it can be done partially, but not completely, because it requires the ability to judge the soundness of a scientific work, which is outside the ability of computers; and, likely, it will be forever.

\section*{Conflict of Interest Statement}
I declare that the research was conducted in the absence of any commercial or financial relationships that could be construed as a potential conflict of interest.

\section*{Author Contributions}
I declare that I am the only author of this work.

\section*{Acknowledgments}
I would like to thank Paola Marziani for inviting me to write this essay. Special thanks also to Marco Berton, Matthew L. Lister, Patrizia Romano, and Merja Tornikoski for a critical reading of an early draft. I also enjoyed an exquisite exchange of words with Robert Antonucci about the anomalous brightness temperatures in jetted AGN on the \href{https://www.facebook.com/groups/activegalacticnuclei}{Facebook group Active Galactic Nuclei}. Thanks also to Beatriz Mingo and Travis Cody Fischer for their suggestions.

\bibliographystyle{ieeetr}
\bibliography{foschini}
\end{document}